# Phase diagram and superlattice structures of monolayer phosphorus carbide ($P_xC_{1-x}$)

Xiaoyang Ma[1], Jun Zhou[1], Tong Yang[1], Dechun Li[2], Yuan Ping Feng[1,3,*]

[1]Department of Physics, National University of Singapore, 2 Science Drive 3, Singapore 117551

[2]School of Information Science and Engineering, Shandong University, 72 Binhai Road, Qingdao 266237, China

[3]Centre for Advanced 2D Materials, National University of Singapore, 6 Science Drive 2, Singapore 117546

*phyfyp@nus.edu.sg

**ABSTRACT**

Phase stability and properties of two-dimensional phosphorus carbide, $P_xC_{1-x}$ ($0 \leq x \leq 1$), are investigated using the first-principles method in combination with cluster expansion and Monte Carlo simulation. Monolayer $P_xC_{1-x}$ is found to be a phase separating system which indicates difficulty in fabricating monolayer $P_xC_{1-x}$ or crystalline $P_xC_{1-x}$ thin films. Nevertheless, a bottom-up design approach is used to determine the stable structures of $P_xC_{1-x}$ of various compositions which turn out to be superlattices consisting of alternating carbon and black phosphorene nanoribbons along the armchair direction. Results of first-principles calculations indicate that once these structures are produced, they are mechanically and thermodynamically stable. All the ordered structures are predicted to be semiconductors, with bandgap (PBE) ranging from 0.2 to 1.2 eV. In addition, the monolayer $P_xC_{1-x}$ are predicted to have high carrier mobility, and high optical absorption in the ultraviolet region which shows a red-shift as the P:C ratio increases. These properties make 2D $P_xC_{1-x}$ promising materials for applications in electronics and optoelectronics.

# I. INTRODUCTION

Alloying is an effective way of tuning properties of materials and it has been used widely in industries. Two or more materials can be mixed uniformly, under certain conditions, to form a solid solution which may have properties intermediate of, and sometimes even beyond, those of the constituting materials. This way of designing new functional two-dimensional (2D) materials has yet been fully explored, even though 2D materials have attracted tremendous interest. Graphene and phosphorene are among the most extensively studied 2D materials. Graphene, a single layer of carbon in the hexagonal structure, is a semimetal, with linear dispersion in its band structure (Dirac cone) near the Fermi level [1]. Black phosphorene, a layer of black phosphorus in a puckered structure, is a semiconductor [2]. One can expect that if carbon and phosphorus can form 2D alloys, they may exhibit interesting physics and exotic properties. Therefore, it would be interesting to find out whether carbon and phosphorus can form 2D phosphorus carbide (PC) alloys, $P_xC_{1-x}$ ($0 \leq x \leq 1$). Given that both C atom in graphene and P atom in phosphorene are three-fold coordinated, it should be possible in principles to form binary PC monolayers.

Earlier effort was made to synthesize PC films. Kuo *et al.* managed to add up to 89% of phosphorus into diamond-like carbon (DLC) films [3]. However, they found that incorporation of high concentration of phosphorus causes amorphization of the film. Nevertheless, the bandgap, as a function of P:C ratio, shows an upward bowing, and the films with low P:C ratios were found suitable for field emission [3]. Pearce *et al.* also doped phosphorus into DLC films up to a P:C ratio of 3:1 [4]. In contrast, they found that the bandgap of the amorphous films decreases linearly with the P contents in the film. It is believed that due to the different bonding preferences of C and P atoms, the stable geometries of PC are likely to suffer from competition between the *sp*$^2$ bonding and *sp*$^3$ bonding [5]. Furlan *et al.* deposited phosphorus-carbide ($CP_x$) thin solid films by unbalanced reactive magnetron sputtering and investigated their properties [6]. They found that $CP_{0.1}$ films exhibit a structure with elements of short-range ordering in the form of curved and inter-locked fullerene-like fragments, but higher deposition temperatures lead to $CP_x$ films with an increasingly amorphous structure. The hardness of the $CP_x$ films was found correlated to the microstructure and a high degree of local ordering (at $x = 1.0$) yields the hardest films (24.4 GPa). More recently, few-layer black-phosphorus carbide *p*-type field effect transistor (p-FET) with high mobility at room temperature was fabricated via a novel carbon doping technique [7]. The same group recently developed a mid-infrared (mid-IR) metasurface by nanostructuring a thin layer of black-phosphorus carbide and realized efficient excitation of hybrid plasmon mode at deep subwavelength-scale [8]. More recently, high-rate high-capacity lithium storage of black-phosphorus composites with graphite was reported [9]. Bulk and nanostructured phosphorus carbides have also drew considerable interest and have been the focus of many studies [10-16].

Theoretical and computational studies on 2D PC have focused mainly on monolayer PC with 1:1 phosphorus to carbon (P:C) ratio [5,17-20]. A number of stable phases have been proposed, and their properties and potential applications have been explored, using mainly first-principles approaches. It is noted that there has been a recent surge in the interest in this group of materials [21-29]. These studies revealed that some of the 2D $P_xC_{1-x}$ have exotic properties such as superconductivity [28], quantum Hall effect [30], high carrier mobility [17], *etc*. A range of potential applications of 2D PCs have been proposed, including anode materials of alkali-metal ion batteries [9,18,20], electrocatalyst for hydrogen evolution reaction [21,23], gas sensors [27], and in high performance nanoelectronics and nano-optoelectronics [17,26]. Compared to PC with 1:1 phosphorus to carbon ratio, a few studies have focused on 2D PCs with other P:C ratios, particularly on monolayer $P_xC_{1-x}$. Two structures (a-$C_2P_4$, $P\bar{4}2\bar{1}m$, and b-$C_2P_4$, $P\bar{1}$) were proposed by Fu *et al.* using the particle swarm optimization (PSO) method [31]. Both were found semiconductors with relatively high electron mobility. Using first-principles calculations and orbital analysis, Huang *et al.* proposed a structure for 2D $P_2C_3$, and found that it possesses double Kagome bands [32]. Very recently, a monolayer structure was proposed for $P_{0.75}C_{0.25}$ ($CP_3$) which was found metallic via first-principles calculations [25,33]. In addition, $P_5C_1$, $P_4C_2$, $P_2C_4$ and $P_1C_5$, resulting from doping of $\beta_0$-PC, were also studied by Rajanshi and Sarkar [19].

These studies suggest the possibility of fabricating monolayer $P_xC_{1-x}$ and their potential technological applications. However, it is obvious that $P_xC_{1-x}$ with arbitrary composition have been left largely unexplored. A systematic study of 2D $P_xC_{1-x}$ would be of both scientific and technological importance. For example, it would be interesting to find out whether it is possible to form 2D $P_xC_{1-x}$ alloy of arbitrary composition (*x*), and if yes, under what conditions? How does the properties of 2D $P_xC_{1-x}$ alloy depend on the composition *x*? In particular, monolayer carbon (graphene) exists in planner honeycomb structure while the most stable monolayer phosphorus, black phosphorene, has a puckered structure. If carbon and phosphorus can form stable 2D $P_xC_{1-x}$, how does the structure of $P_xC_{1-x}$ evolve with the composition? What would be the phase diagram of 2D $P_xC_{1-x}$?

Motivated by these questions, we first carried out a systematic investigation on possible formation of monolayer $P_xC_{1-x}$ using the first-principles method in combination with cluster expansion (CE) and Monte Carlo (MC) simulation. It is found, unfortunately, that monolayer $P_xC_{1-x}$ is a phase separating system, and equilibrium growth of PC would lead to separate C-rich and P-rich phases. Nevertheless, we proceed with design of 2D $P_xC_{1-x}$ alloys by taking a bottom-up approach, and studied the stability and properties of the most stable structure at $x = 0.125, 0.25, 0.375, 0.5, 0.625, 0.75$ and $0.875$. Even though 2D $P_xC_{1-x}$ is predicted to phase separate, such structures could be fabricated using non-equilibrium methods or stabilized by a substrate. Our study reveals that the most stable structure at each composition considered is of the form of a superlattice, consisting of alternating carbon and phosphorus nanoribbons along the armchair direction. All monolayer $P_xC_{1-x}$ are semiconductors with bandgap (PBE) ranging from 0.2 eV to 1.2 eV. Moreover, the ordered $P_xC_{1-x}$ alloys are predicted to have high mobility, reaching up to $10^4$ cm$^2 \cdot $V$^{-1} \cdot$s$^{-1}$, which is comparable to that of graphene. The high mobility and tunable bandgap are desired for many applications such as electronics, optoelectronics and energy storage.

## II. METHODS

Stabilities of the 2D $P_xC_{1-x}$ alloys are investigated using the CE method which has been widely used to predict the energetics of alloys. The CE implemented in the MIT *ab-initio* phase stability (MAPS) package in Alloy Theoretic Automated Toolkit (ATAT) [34,35] is used in the present study. Both stable 2D structures of carbon and phosphorus are considered for parent structures in the simulation. While graphene is the only stable structure for 2D carbon, there exist several relatively stable $sp^3$ bonded phosphorene allotropes [36], such as black-, blue- and γ-phosphorene (see Fig. S1 and Table S1 in the Supplement Materials (SM) [37]). Therefore, CE calculations are carried out using the planner structure of graphene and black-, blue- and γ-phosphorene as the parent structure, respectively.

Given the lattice information of the parent structure, increasingly more accurate CEs can be constructed by including larger and larger clusters. In the CE scheme, occupation of an atomic site ($i$) in a cluster is represented by an occupation variable, $\sigma_i$, which takes the value of $-1$ or $+1$ depending on the type of atom occupying the site, and an atomic configuration $\sigma$ contains the value of the occupation variable for each site in the parent lattice. The average energy per atom, $E(\sigma)$, of the alloy is given as a polynomial in the occupation variables [34],

$$E(\sigma) = \sum_\alpha m_\alpha J_\alpha \langle \prod_{i \in \alpha'} \sigma_i \rangle \quad (1)$$

where $\alpha$ is a cluster, $J_\alpha$ is the effective cluster interaction (ECI), and the multiplicity factor $m_\alpha$ indicates the number of clusters that are equivalent by symmetry to $\alpha$. The precision of the ECIs can be evaluated by the cross-validation score, defined as

$$CV = \frac{1}{n} \sum_{i=1}^{n} \left( E_i - \hat{E}_{(i)} \right)^2 \quad (2)$$

where $n$ is the number of structures used to obtain the ECIs, $E_i$ and $\hat{E}_{(i)}$ denote the energy calculated using the density functional theory (DFT) and the predicted value of the energy of structure $i$, respectively. To ensure a satisfactory convergence of CEs, all the cross-validation scores are less than 0.025 eV. With the appropriate ECI, the thermodynamic potential of each phase can be obtained by performing the semi-grand-canonical MC simulation. The phase boundary which is given by the intersection of the thermodynamic potentials of two phases in equilibrium is determined by running two MC simulations simultaneously, one for each phase, at temperature and concentration grids of $\Delta T = 10$ K and $\Delta x = 0.001$, respectively. The simulation cell, represented by an equivalent circle of 50 Å, containing more than 2000 atoms, is used in the MC simulation [34,35,38].

In the DFT calculations, the projector augmented-wave (PAW) method and the Perdew-Burke-Ernzerhof form of generalized gradient approximation (GGA-PBE) for exchange-correlation interaction [39] implemented in the Vienna *ab initio* simulation package (VASP) [40] are adopted. The cut-off energy of planewave expansion of electron wavefunction was set to 600 eV. The convergence was confirmed when the difference in total energy was less than $10^{-6}$ eV and residual force on each atom was less than 0.01 eV Å$^{-1}$. The spacing between consecutive k points was less than 0.03 Å$^{-1}$. The relative stability of the 2D structures is assessed based on the calculated formation energy which is defined as

$$E_f(x) = E_{P_xC_{1-x}} - xE_P - (1-x)E_C \tag{3}$$

where $E_{P_xC_{1-x}}$, $E_C$ and $E_P$ and are the total energy per atom of $P_xC_{1-x}$, graphene, and the concerned phosphorene, respectively.

According to the deformation potential theory of Bardeen and Shockley, the carrier mobility based on the phonon-limited scattering model of a 2D material can be obtained by [41-43],

$$\mu^{2D} = \frac{e\hbar^3 C^{2D}}{k_B T m^* m_d E_1^2} \tag{4}$$

where $T$ is the temperature, $e$ the electron charge, $\hbar$ and $k_B$ are the Plank and Boltzmann constant, respectively. $m^*$ is the effective mass of electron and hole in the transport direction, $m_d = \sqrt{m_x^* m_y^*}$ is the average effective mass. The carrier effective mass is obtained from the second derivative at the extremum of the conduction or valence band. $C^{2D}$ in Eq. (4) is the in-plane stiffness constant, and $E_1$ is the deformation potential which can be calculated from the change in band energy $\Delta E$ for a given change of $\Delta l$ in the lattice constant $l_0$, as $E_1 = \Delta E/(\Delta l/l_0)$.

To access the optical properties of the materials, the complex dielectric function $\varepsilon(\omega) = \varepsilon_1(\omega) + i\varepsilon_2(\omega)$, where $\omega$ is the photon frequency, is calculated. Based on the Maxwell equations, the absorption of a 2D material is related to the in-plane 2D optical conductivity, $\sigma_{2D}(\omega)$ [44-46],

$$\sigma_{2D}(\omega) = iL[1 - \varepsilon(\omega)]\varepsilon_0 \omega \tag{5}$$

where $L$ and $\varepsilon_0$ are the slab thickness and the permittivity of vacuum, respectively. The dielectric function $\varepsilon(\omega)$ can be expressed in the Ehrenreich-Cohen formula, in which $\varepsilon(\omega)$ is directly related to the eigenstates $|c\mathbf{k}\rangle$, $|v\mathbf{k}\rangle$ and eigenvalues $\varepsilon_c(\mathbf{k}), \varepsilon_v(\mathbf{k})$ of the conduction (c) and valence (v) bands [47], respectively.

$$\varepsilon(\omega) = 1 + \frac{2e^2\hbar^2}{\varepsilon_0 m^2} \frac{1}{LS} \sum_{c,v} \sum_{\mathbf{k}} \frac{|\langle c\mathbf{k}|p_j|v\mathbf{k}\rangle|^2}{[\varepsilon_c(\mathbf{k}) - \varepsilon_v(\mathbf{k})]^2} \times \sum_{\beta=+,-} \frac{1}{\varepsilon_c(\mathbf{k}) - \varepsilon_v(\mathbf{k}) - \beta(\hbar\omega + i\eta)} \tag{6}$$

where $S$ and $\mathbf{k}$ are the area of the unit-cell and the 2D wave-vector, respectively. $p_j$ represents the $j$th component of the momentum operator, and $\eta$ takes the value of +1 (-1) for the conduction (valence) band. It is noted that $\sigma_{2D}(\omega)$ is independent of $L$. The absorbance $A(\omega)$ at normal incidence is then given by [44-46]:

$$A(\omega) = \frac{\mathrm{Re}\sigma_{2D}(\omega)}{\varepsilon_0 c} \tag{7}$$

where Re($z$) represents the real part of $z$ and $c$ is the speed of light.

Electronic structures obtained from DFT calculations with the GGA-PBE functional are used in calculating the optical properties. Even though GGA-PBE suffers from problems such as underestimation of bandgap of semiconductors, and the optical properties of 2D materials are also strongly affected by both intense quasi-particle correction on the electronic band structure and strong excitonic effects, results obtained from the simple GGA-PBE calculations are reliable in describing the trend in optical properties of related materials and are useful when the optical properties of different materials are compared which is our interest here. Depending on the structures, 120 to 180 empty bands (3 times of the default settings in VASP) are included in the calculations of optical properties. Test calculations with different empty bands, up to 240, were performed and it was found that the chosen value is sufficient to produce reliable results.

### III. RESULTS AND DISCUSSION

#### A. Phase Diagram of Monolayer $P_xC_{1-x}$ Alloys

We first explore the feasibility of forming 2D $P_xC_{1-x}$ ($0 \leq x \leq 1$) alloys, using the first-principles method in combination with cluster expansion and Monte Carlo simulation. The monolayer structure of graphene, black-phosphorene, blue-phosphorene, and γ-phosphorene, which are shown in Fig. S1 in the SM [37], is used as the parent structure, respectively, in the CE simulation. The obtained energy-composition ($x$) diagrams of $P_xC_{1-x}$ are presented in Fig. 1. The energies predicted by the CE method are superimposed with those from the first-principles calculations based on the DFT. The energy presented in the figure is the formation energy calculated using Eq. (3), with $E_P$ is taken to be the energy of the concerned parent structure (graphene or one of the phosphorene phases) in each case. To ensure a satisfactory convergence of CE, 83, 116, 105 and 81 structures are used for fitting while 1269, 1669, 847 and 969 structures are used for prediction in Figs. 1(a), 1(b), 1(c) and 1(d), respectively, and the cross-validation scores are 8.0, 9.8, 10.0 and 8.5 meV/atom, respectively.

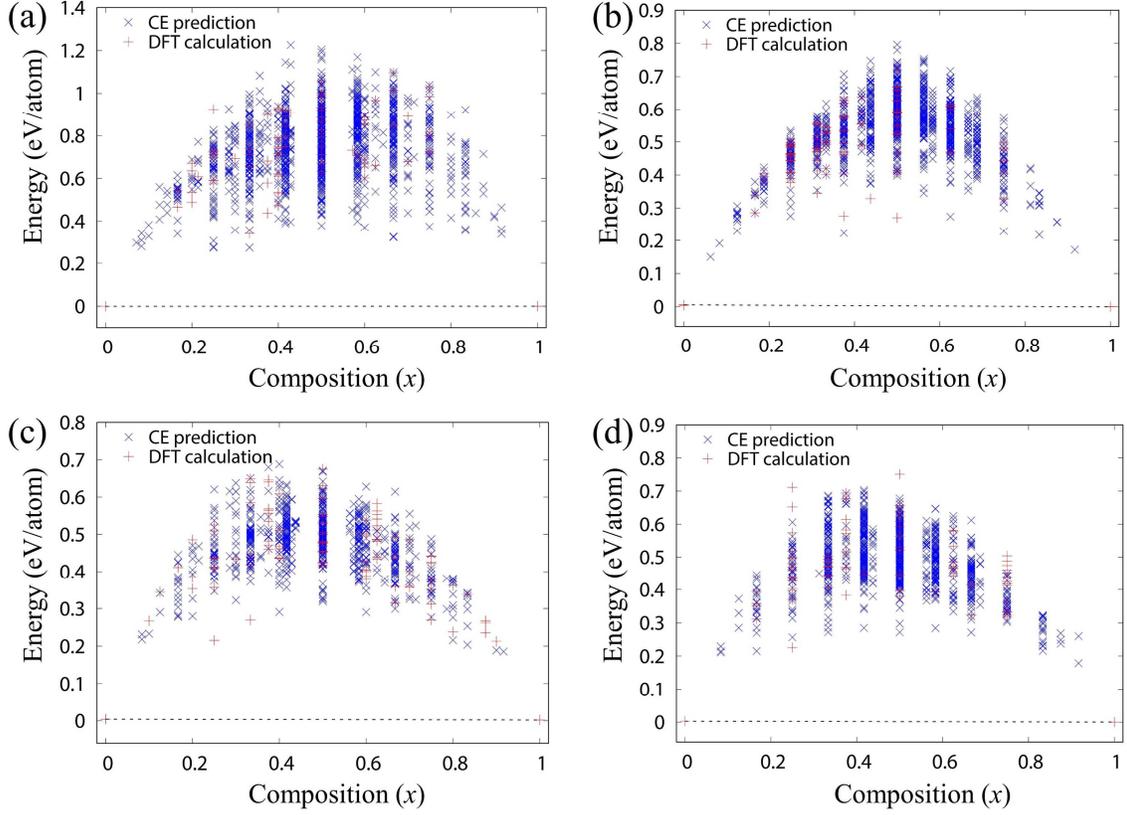

FIG. 1. Energy of monolayer $P_xC_{1-x}$ obtained using (a) graphene, (b) black-phosphorene, (c) blue-phosphorene and (d) γ-phosphorene as the parent structure, respectively. The energy is calculated using Eq. (3), with $E_P$ taken to be the energy of the corresponding parent phosphorene phase in (c)-(d). The blue × and red + indicate CE-predicted energy and the DFT-calculated energy, respectively. 83, 116, 105 and 81 structures are used for fitting while 1269, 1669, 847 and 969 structures are used for prediction in Figs. 1(a), 1(b), 1(c) and 1(d), respectively, and the cross-validation scores are 8.0, 9.8, 10.0 and 8.5 meV/atom, respectively.

It is clear that the formation energies of all configurations in the whole composition range are at least 100 meV/atom. This suggests that forming the binary $P_xC_{1-x}$ alloy is an endothermic process, and separate phases of graphene and phosphorene are energetically favoured compared to a binary PC alloy. These results explain the difficulty in fabricating 2D $P_xC_{1-x}$ or crystalline PC thin films [3,4,6].

To further evaluate the energetics of the $P_xC_{1-x}$ alloys, we calculated the phase diagram of monolayer $P_xC_{1-x}$ and show the result in Fig. 2. The results confirm that $P_xC_{1-x}$ prefers phase separation, rather than forming solid solution, regardless of the parent structure. The phase separation region extends to very high temperature and does not show any tendency of mixing except for very small or large $x$. Only the phase diagram for $T \geq 6000$ K is shown in Fig. 2 because the high critical temperature is beyond the reach of any experiment. The phase diagram is also not corrected by the vibration contribution which may affect the order-disorder transition temperature [48]. However, considering the very high critical temperatures of $P_xC_{1-x}$, the correction by vibrational contribution would be marginal and would not lead to any significant changes to the phase diagram. The small possibility of mixing at high temperature and small or large $x$ is similar to doping of P in graphene or C in phosphorene. Even though the monolayer structure is different from that of DLC films, the doping effect would be similar. The low solubility of P in graphene found here is in qualitative agreement with experimental results reported by Kuo *et al*. who found that when more than 5% of P is doped to DLC films, the P atoms gradually destroy the long-medium range structure associated with C-C bonding [3]. In addition, asymmetric phase boundaries at low and high $x$ are observed in Fig. 2, particularly for the structures obtained using γ-P as the parent structure. This is due to the different solubilities of C in γ-P and P in γ phase carbon. The phase diagram shown in Fig. 2 nevertheless suggests a low solubility of P in graphene or vice versa, and the difficulty in fabricating uniform monolayer $P_xC_{1-x}$ binary alloys using equilibrium growth methods.

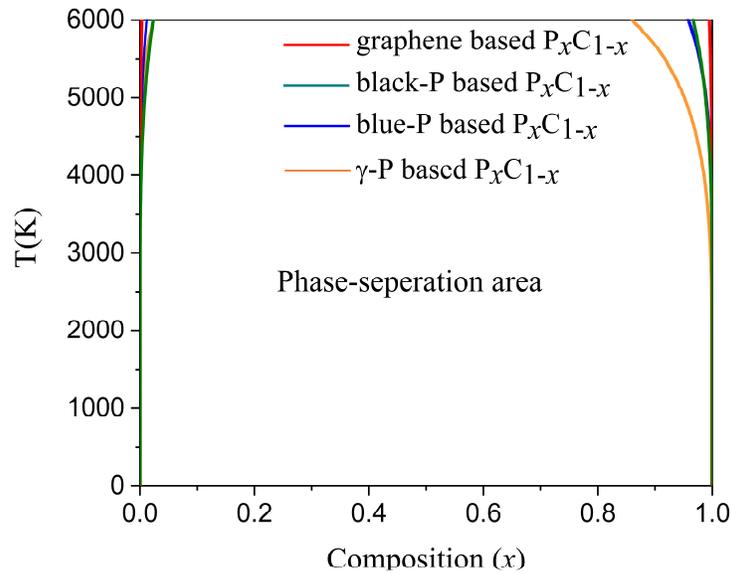

FIG. 2. Phase diagram of $P_xC_{1-x}$ obtained using graphene, black-, blue- and γ-phosphorene as the parent structure, respectively. The 2D binary system is predicted to phase separate for the entire composition range, except for very C-rich or very P-rich region at very high temperature.

### B. Bottom-up Design of $P_xC_{1-x}$ Alloys

In order to understand why $P_xC_{1-x}$ prefer phase separation and to further explore possible monolayer $P_xC_{1-x}$ structures, we carried out a systematic bottom-up design of monolayer $P_xC_{1-x}$. For each composition ($x$), we consider all possible atomic arrangements in a supercell containing up to 16 atoms. High-throughput first-principles calculations are then carried out to optimize each unique structure and evaluate its stability. Even though our CE simulation predicts $P_xC_{1-x}$ to phase separate, such a bottom-up design is meaningful, because metastable $P_xC_{1-x}$ binary alloys could be grown using non-equilibrium methods, as demonstrated in recent experiments [3,7]. Moreover, 2D materials can often be stabilized by a substrate, even though they may be unstable in the suspended form which is the case in our CE simulation.

The bottom-up design process described above, using each of the four graphene/phosphorene phase (graphene, black-, blue- and γ-phosphorene) in turn as the parent 2D lattice structure, resulted in 2282 unique configurations of $P_xC_{1-x}$, after discarding some structures which have very high formation energy. High-throughput DFT calculations were performed to calculate the formation energy, using Eq. (3), with $E_P$ taken to be the energy of black phosphorene, the most stable phase of phosphorene. The results are presented in Fig. 3, where the lowest energy for each composition is indicated by a red-colored square. As expected, the overall distribution of the energy values is similar to that shown in Fig. 1. All configurations have positive formation energy, implying that $P_xC_{1-x}$ are metastable compared with graphene and phosphorene.

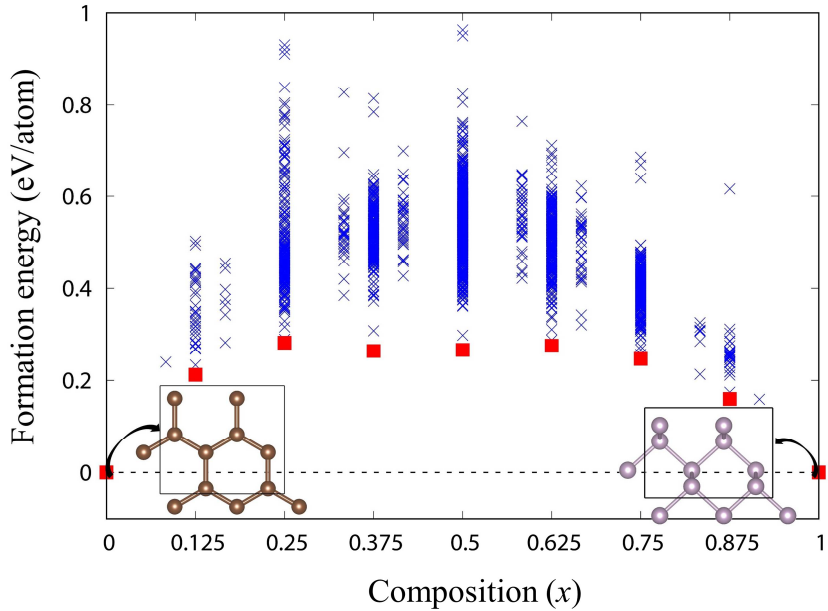

FIG. 3. DFT calculated formation energy (per atom) of $P_xC_{1-x}$ configurations generated using the bottom-up approach. The lowest energy for each composition ($x$) is indicated by the red square.

The structure corresponding to the minimum energy in Fig. 3 for each composition considered is shown in Fig. 4. It is obvious that all these are phase separating structures, in which the carbon atoms and the phosphorus atoms form separate one-dimensional clusters. The carbon area maintains essentially the flat hexagonal graphene structure, while the structure of the phosphorus area has resemblance to the puckered black phosphorene. In C-rich structures, the P atoms act as a bridge connecting the $sp^2$ bonded C clusters. Similarly, in P-rich structures, the C atoms link the $sp^3$ bonded P clusters. This is because carbon prefers $sp^2$ bonding but phosphorus energetically favors $sp^3$ bonding, and forming separate carbon and phosphorus clusters reduces competition between the two types of bonding and the energy of the system. Another interesting common feature in these structures is the alternating carbon and phosphorus linear strips (nanoribbons) along the armchair direction. It is noted that the interface between the carbon area and the phosphorus area is always along the armchair direction. This is because of the better lattice matching between graphene and phosphorene in the armchair direction (4.275 Å of graphene vs. 4.629 Å for black phosphorene), compared to that in the zigzag direction (2.468 Å vs. 3.300 Å). Due to the small lattice mismatch in the armchair direction, there exists a small tensile/compressive strain (a few percent) along the graphene/phosphorene nanoribbon which leads to some small but insignificant changes in bond lengths and bond angles in the graphene and phosphorene nanoribbons. Overall, the structures of the C and P nanoribbons are very similar to slightly strained graphene and black phosphorene nanoribbons, except the distortions at the edges. Therefore, strain effect is not expected to play an important role in determining the electronic properties of the superlattice structures, compared to other effect such as quantum confinement and edge/interface effect. Even though the constraint of the finite supercell size may play a role in the formation of such patterned structures, the results do suggest easy formation of graphene-black-phosphorene superlattices. Such structures provide 1D carbon and phosphorus channels and may lead to anisotropy of $P_xC_{1-x}$ (see discussion below).

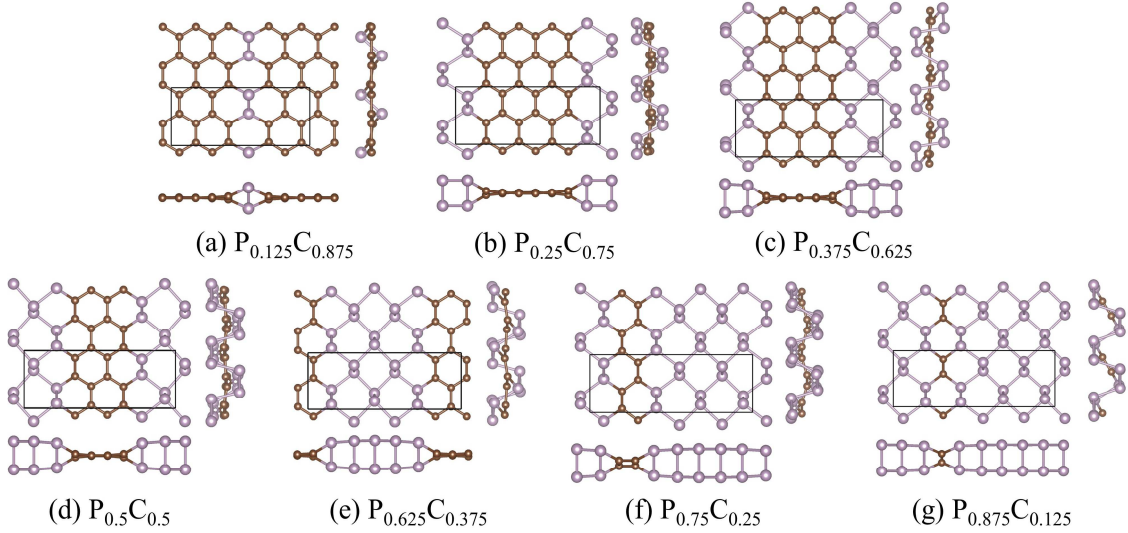

FIG. 4. Relaxed minimum energy structures of $P_xC_{1-x}$ for (a) $x = 0.125$, (b) 0.25, (c) 0.375, (d) 0.5, (e) 0.625, (f) 0.75, and (g) 0.875, obtained from the bottom-up approach. The C and P atoms are shown in brown and purple colors, respectively. The rectangle indicates the supercell used to generate the structure. A top view and two side views are presented in each case. The corresponding lattice parameters are given in Table S2 in SM [37].

We calculated phonon spectra of these structures and performed *ab initio* molecular dynamics (AIMD) simulations to evaluate their stability. The results presented in Fig. S2 in the SM[37] confirm that once fabricated, these structures are both dynamically and thermodynamically stable, even at room temperature. This is expected since the superlattice structure has the lowest energy among all structures generated for each composition within the constraint of the finite size of the supercell. It can be expected that normal growth processes such as CVD would produce randomly distributed segregated phases of P and C, similar to segregated domains of boron nitride and carbon in 2D hBN-C which was demonstrated to be a phase separating 2D material in recent studies [49-51]. However, with some control, the superlattice structures shown in Fig. 4 may be fabricated. By varying the P:C ratio and the superlattice patterns, the physical properties of these ordered structures can be tuned to realize bandgap-engineered applications in electronics and optics (see Section C).

Among these 2D $P_xC_{1-x}$ alloys, $P_{0.5}C_{0.5}$ with 1:1 phosphorus to carbon ratio has been most well studied. However, smaller supercells were used in previous studies and only highly ordered structures were obtained [7,17-20]. Compared to these highly ordered structures which are included in the present study, the structure identified as the most stable one in the present work (Fig. 4(d)) has lower energy and is predicted to be more stable. Similar to the stable structures obtained for other compositions, $P_{0.5}C_{0.5}$ phase separates into carbon and phosphorus nanoribbons along the armchair direction. The carbon and phosphorus areas roughly maintain the structure of graphene and black phosphorene, respectively, but significant distortions are observed at the interface. In addition to room temperature, AIMD simulation is also carried out at a high temperature of 800 K, using a 64-atom supercell (see inset in Fig. S2(d) in SM[37]). As shown in Fig. S3(b) in the SM[37], the atomic structure of $P_{0.5}C_{0.5}$ is very well maintained during the AIMD simulation up to 18 ps. The variation in the total energy is within 0.04 eV/atom. The simulation results confirm the stability of $P_{0.5}C_{0.5}$.

The electronic configuration around the C and P atoms are also similar to those in graphene and black phosphorene, respectively. The calculated electron localized function (see Fig. S3(a) in SM[37]) confirms formation of covalent P-C bond at the interface. The calculated band structure (see Fig. S2(d) in SM[37]) indicates that $P_{0.5}C_{0.5}$ is a semiconductor with an indirect bandgap of 1.12 eV. The valence band maximum (VBM) and the conduction band minimum (CBM) are located at point X (0.5, 0) and along line D that connects point Γ (0,0) and point Y (0, 0.5), respectively, in the first Brillion zone of the reciprocal lattice. These bands are dominated by the *p* orbitals of both C atoms and P atoms. The direct bandgap, located at Y, is very close to the indirect bandgap in energy and the calculated value (GGA-PBE) is 1.2 eV.

Even though the crystal structure of $P_{0.5}C_{0.5}$ consists of alternating graphene and black-phosphorene nanoribbons along the armchair direction, this monolayer phase shows almost isotropic properties. The in-plane sound velocity, derived from the calculated phonon spectrum, is 9.4 km·s$^{-1}$ along the armchair direction and 9.6 km·s$^{-1}$ in the zigzag direction. The close values indicate an isotropic in-plane stiffness of the 2D crystal. The calculated electron effective masses in the armchair direction (0.36 $m_0$) and zigzag direction (0.44 $m_0$) are also fairly close, while the electron mobility in the zigzag direction (7.29 × 10$^3$ cm$^2$V$^{-1}$s$^{-1}$) is about two times of that along the armchair direction (3.48 × 10$^3$ cm$^2$V$^{-1}$s$^{-1}$). For holes, the effective masses are 0.33 $m_0$ and 0.44 $m_0$, and the mobilities are 1.56 × 10$^3$ cm$^2$V$^{-1}$s$^{-1}$ and 1.44 × 10$^3$ cm$^2$V$^{-1}$s$^{-1}$ along the armchair and the zigzag direction, respectively. This is in sharp contrast to the anisotropic properties reported earlier for 2D PC [17]. Using DFT-PBE calculation, the authors in Ref. [17] predicated order-of-magnitudes different mobilities in the zigzag and armchair directions for both electron and hole for their α- and β-phase of 2D PC. This discrepancy is most likely due to the details of atomic arrangement. While the $P_{0.5}C_{0.5}$ superstructure obtained in the present study consists of graphene and black phosphorene nanoribbons, both the α- and β-phases of 2D PC in Ref. [17] consist of checker board arrangement of C and P dimers in the armchair direction. While it is interesting to further investigate the transport properties of these 2D structures, our calculations indicate that these checker board structures are unstable compared to the nanoribbon superstructure obtained in the present study and therefore it would be much more difficult, if possible, to realize them experimentally. Isotropy is also observed in the calculated mechanical properties of $P_{0.5}C_{0.5}$ (see calculated Young's modulus and Poisson's ratio in Table S4 in SM[37]). Besides this most stable structure (Fig. 4(d)) and those reported in previous studies, we also obtained three other structures for $P_{0.5}C_{0.5}$ with good stability (Fig. S4 in SM[37]). Their formation energy is 0.12, 0.19 and 0.27 eV/atom, respectively, above that of the most stable phase.

It is interesting to note that when one in every eight carbon atoms in the planner graphene structure is replaced by P, the semi-metallic graphene is turned into a semiconductor with a significant bandgap of 0.83 eV (Fig. S2(a) in SM[37]). This is related to the unique superlattice structure, as discussed in the next subsection. In the structure optimized using the 16-atom supercell, the two P atoms form a buckled dimer which leads to a P dimer row along the armchair direction, linking armchair graphene nanoribbons of 7 C-chain wide in the stable phase of $P_{0.125}C_{0.875}$. As a result, each P atom is $sp^3$ bonded with its neighboring P atom and two C atoms, while the C atoms remain in the planner structure, even though the C-P bonds at the interface are tilted (Fig. 4(a)). The stability of this $P_{0.125}C_{0.875}$ phase is confirmed by the calculated phonon spectrum and AIMD simulation at room temperature and 800 K (Fig. S5 and Videos S1 and S2 in SM[37]). As revealed by the projected density of states shown in Fig. S6 in SM[37], the highest valence band of $P_{0.125}C_{0.875}$ consists of mainly the $p_y$ orbital of the two P atoms, and the $p_y$ orbital of the C atoms at the interface, while the lowest conduction band is dominated by the $p_y$ orbital of the interface C atoms.

Compared to $P_{0.5}C_{0.5}$, $P_{0.125}C_{0.875}$ exhibits more anisotropy in its physical properties. The in-plane speed of sounds derived from the phonon spectrum is 21.5 and 17.6 km·s$^{-1}$ along the armchair and zigzag direction, respectively. Significant anisotropy is also seen in the calculated carrier mobility. Even though the calculated electron effective mass in the armchair (0.23 $m_0$) and the zigzag (0.22 $m_0$) directions are close, the electron mobility in the zigzag direction (18.1 × 10$^3$ cm$^2$V$^{-1}$s$^{-1}$) is 9 times higher than that in the armchair direction (2.02 × 10$^3$

cm$^2$V$^{-1}$s$^{-1}$). The transport property of holes displays more disparity along the two crystal directions, with effective mass of $0.17\,m_0$ and $0.84\,m_0$ and mobility of $38.5 \times 10^3$ cm$^2$V$^{-1}$s$^{-1}$ and $0.38 \times 10^3$ cm$^2$V$^{-1}$s$^{-1}$ along the armchair and zigzag direction, respectively.

Two additional metastable structures, shown in Fig. S7 in SM[37], are found for $P_{0.125}C_{0.875}$. Both of them emerged during structural optimization of ML structures of $P_{0.125}C_{0.875}$. Their formation energies are 0.23 and 0.31 eV/atom, respectively, above that of the most stable structure in Fig. 4(a). The first one (see Fig. S7(a) in SM[37]) is similar to the most stable structure (Fig. 4(a)), with P dimer row linking the armchair C nanoribbons. However, the P dimer is parallel to the graphene plane. To form $sp^3$ bond which is energetically favoured by P atom, the graphene nanoribbon is bent. The second structure shown in SM, Fig. S7(b) [37] also consists of P rows between C nanoribbons. However, this is a bilayer structure, in which the interface C atom forms an additional bond with another C atom in the opposite layer, and each P atom is covalently bonded to three C atoms. Carbon atoms in the nanoribbon maintain the planner honeycomb structure. The $P_{0.125}C_{0.875}$ structure shown in Fig. S7(a) is a semiconductor, with a direct bandgap (GGA-PBE) of 0.64 eV, while the one shown in Fig. S7(b) is metallic. The dynamic stabilities of both structures are confirmed by the calculated phonon spectrum (Fig. S7 in SM[37]) which exhibit negligible imaginary frequency. However, during the AIMD simulation at room temperature (300 K), a phase transition from the structure shown in Fig. S7(a) in SM[37] to the most stable structure (Fig. 4a) is detected (see Video S3 in SM[37]).

### C. Composition Dependent Properties of 2D $P_xC_{1-x}$ Superlattices

Finally, we present composition dependent properties of the 2D $P_xC_{1-x}$ superlattices along with those of graphene $(x = 0)$ and black-phosphorene $(x = 1)$. First of all, all 2D $P_xC_{1-x}$ superlattices, except graphene, are semiconductors. The calculated bandgap is shown in Fig. 5(a) against composition ($x$). Bandgap values from previous studies are also included in the figure for comparison [5,7,17,19-23,25,28,29,31-33]. It is clear that the bandgap obtained in the present study fluctuates with the P:C ratio but they are within the range of values obtained in previous studies. As shown by the calculated PDOS presented in Fig. S2 in SM, the electronic states at the valence and conduction band edges are contributed by both carbon and phosphorous orbitals. The electronic properties of the superlattice structures may have some resemblance to those of graphene/phosphorene nanoribbons but would be significantly modified due to the presence of different neighboring nanoribbons and the interface effect. Nevertheless, at low P:C ratios, the band edges are dominated by carbon p orbitals. In this case, one would expect that the electronic properties would be similar to that of armchair graphene nanoribbons (AGNRs). It is known that the bandgap of an AGNR is direct and oscillates with the width of the nanoribbon. In particular, the bandgaps of AGRRs consisting of $2p + 1$ ($2p + 2$), where $p$ is an integer, carbon chains is large (small), while those with $3p$ carbon chains have intermediate bandgap values [52]. It is noted that the bandgap variation of $P_{0.152}C_{0.875}$ (equivalent to AGNRs of width 7), $P_{0.25}C_{0.75}$ (width 6), $P_{0.375}C_{0.625}$ (width 5), $P_{0.5}C_{0.5}$ (width 4), and to some extent even $P_{0.625}C_{0.375}$ (width 3), follows this trend. In view of the LDA direct bandgap of ~1.6 eV for AGNR of width 7 obtained in Ref. [52] a relatively large direct bandgap can be expected for $P_{0.152}C_{0.875}$ and the value of 0.83 eV obtained in the present study is thus not surprising. At higher P:C ratios, the phosphorous $p$ orbitals play a more important role. The bandgap of armchair phosphorene nanoribbon generally decreases with increasing width of the nanoribbon [53-55]. The calculated bandgaps of $P_{0.75}C_{0.25}$ and $P_{0.875}C_{0.125}$ follow this trend. It is also noted that the bandgaps of $P_{0.125}C_{0.875}$, $P_{0.25}C_{0.75}$, and $P_{0.625}C_{0.375}$ are direct while other $P_xC_{1-x}$ have an indirect gap. The direct gap of $P_{0.625}C_{0.375}$ is at the $\Gamma$ point while that of $P_{0.125}C_{0.875}$ and $P_{0.25}C_{0.75}$ occurs at X in the first Brillion zone in the reciprocal lattice. For the indirect bandgap semiconductors, the VBM of $P_{0.75}C_{0.25}$ is located at the $\Gamma$ point while that of the others ($P_{0.375}C_{0.625}$, $P_{0.5}C_{0.5}$ and $P_{0.875}C_{0.125}$) is at the X point, and the CBM of $P_{0.375}C_{0.625}$ is at the $\Gamma$ point and that of the others ($P_{0.5}C_{0.5}$, $P_{0.75}C_{0.25}$ and $P_{0.875}C_{0.125}$) is along line D ($\Gamma$-Y). The direct/indirect bandgap transition is likely induced by strain along the armchair direction in the superlattice structures. The graphene and black phosphorene regions in the structure can be easily relaxed in the zigzag direction. But in the armchair direction, the lattice constants of graphene and black phosphorene compete and compromise to a value which deviates from the Vegard's law. More importantly, this leads to strain in the armchair direction in both the graphene and black phosphorene nanoribbons. Earlier study predicted that a strain can induce a direct to indirect bandgap transition in edge passivated armchair graphene nanoribbon [56]. Similarly, a strain applied along the armchair direction can lead to a direct to indirect bandgap transition in 2D phosphorene [57]. The observed direct to indirect bandgap transition in the $P_xC_{1-x}$ superlattice structures is likely due to a similar strain effect, even though the exact critical strain may depend on the interface effect and the mixed contributions of C and P orbitals to the band edges, in addition to the ribbon width. This, however, does not explain the exception at $x = 0.625$. The electronic properties of $P_xC_{1-x}$ deserve further investigation. The calculated

PBE bandgap varies between 0 and 1.2 eV in the entire composition range, providing opportunity for bandgap engineering to cater $P_xC_{1-x}$ for requirement of different applications.

Due to the superlattice structures, the calculated phonon dispersion shows a high frequency region (~20-40 THz) dominated by vibration of C-C bonds, and a low frequency region (<20 THz) which is contributed by both carbon and phosphorus. Phonon bandgaps are observed in P-rich monolayer $P_xC_{1-x}$ which could have a significant effect on their thermal transport property. Previous studies on 2D $MoS_2$ and $WS_2$ suggest that phonon bandgap in these 2D materials leads to an increase in their thermal conductivity [58].

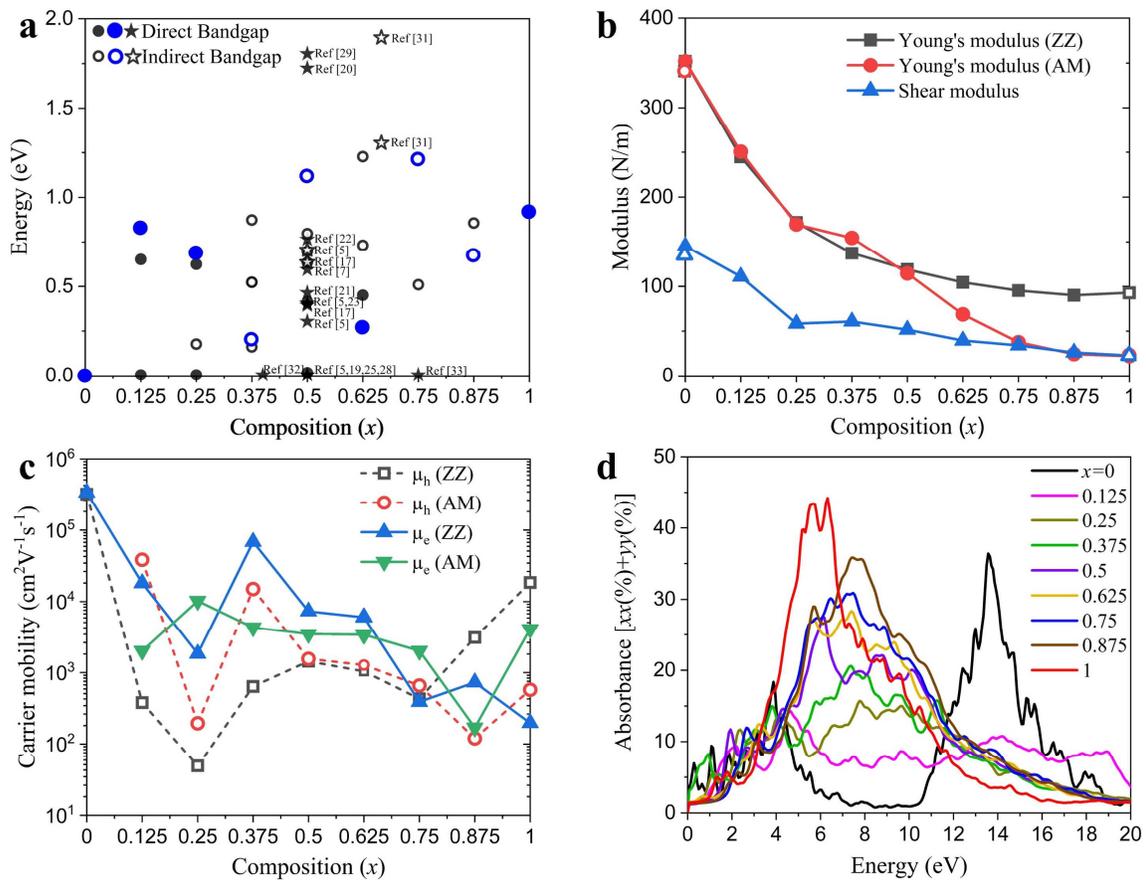

FIG. 5. Composition dependent properties of $P_xC_{1-x}$. Variation of (a) bandgap, (b) Young's moduli and shear modulus, and (c) carrier mobilities with composition. (d) Total absorbance of $P_xC_{1-x}$ of various compositions. Bandgap values from various previous studies are included in Fig (a) for comparison. The reference which reports the bandgap value is indicated next to the data symbol. The open (closed) circles in (a) indicate indirect (direct) bandgap. The values of modulus of graphene and black-phosphorene are close to the values reported earlier in Refs. [59] and [60], respectively, which are shown by the closed symbols in (b).

The mechanical properties of the low energy structures (Fig. 4) are also investigated. The elastic constants are derived from the second-order derivative of the energy versus strain curve [61]. By allowing the strain to vary from -0.015 to 0.015 at an interval of 0.005, the relevant stiffness tensor elements $C_{11}$, $C_{22}$, $C_{12}$, $C_{66}$, the Young's moduli $Y_x$ and $Y_y$, the shear modulus $S_y$, and the Poisson's ratios $\nu_x$ and $\nu_y$ are calculated and the results are presented in Table S4 in SM[37]. We also list the mechanically stability of the structure evaluated based on the calculated elastic properties. The Young's moduli and shear modulus are also shown in Fig. 5(b) as functions of composition. Generally, both Young's moduli and the shear modulus decrease as the P content in monolayer $P_xC_{1-x}$ increases. This is expected because the non-planner structures of phosphorene are more susceptible to deformation by an in-plane strain. It is also interesting to note that at low P:C ratio, where elastic property is mainly determined by graphene which is mechanically isotropic, the stiffness in the armchair and zigzag directions are very close, indicating a mechanically isotropic $P_xC_{1-x}$. However, in P-rich monolayer $P_xC_{1-x}$, the Young's modulus in the armchair direction is much smaller than that in the zigzag direction, implying a mechanical anisotropy similar to that in black phosphorene [62]. This is because strain along the armchair direction can be easily partially absorbed by altering the dihedral angle, while it is mainly compensated by P-P bond elongation along the zigzag direction. This evolution from isotropic C-rich $P_xC_{1-x}$ to anisotropic P-rich $P_xC_{1-x}$ is also reflected in the calculated Poisson's ratios (see Table S4 in SM[37]).

To assess the transport properties of the monolayer $P_xC_{1-x}$, we calculated the effective mass, and carrier mobility at room temperature by Eq. (4) and present the results in Table S3 in SM [37]. The calculated electron and hole mobilities are also shown in Fig. 5(c). Both graphene and phosphorene have high carrier mobilities. The calculated electron and hole mobilities of $P_xC_{1-x}$, in both the armchair and zigzag directions, are comparable to those of graphene and phosphorene, i.e. $\sim 10^3 - 10^4$ cm$^{-2}$V$^{-1}$s$^{-1}$, with a few exceptions, such as the hole mobility of $P_{0.25}C_{0.75}$, which is below 200 cm$^{-2}$V$^{-1}$s$^{-1}$ in the armchair direction and < 100 cm$^{-2}$V$^{-1}$s$^{-1}$ in the zigzag direction. The highest electron mobility (68470 cm$^{-2}$V$^{-1}$s$^{-1}$) is obtained along the zigzag direction of $P_{0.375}C_{0.625}$ while the highest hole mobility is found along the armchair direction in $P_{0.125}C_{0.875}$. Overall, the mobility values decrease slightly from C-rich to P-rich $P_xC_{1-x}$.

It is well-known that PBE functional underestimates bandgaps of semiconductors, and the optical properties of 2D materials are strongly affected by both intense quasi-particle correction on the electronic band structure as well as strong excitonic effects. On the other hand, compared to the state-of-art systematic GW+BSE calculations which is challenging for the superlattice structures, calculation based on PBE is much more efficient and is able to correctly predict the trends in optical properties and thus provides valuable information when optical properties of different materials are compared which is our interest here. The PBE-calculated optical absorbance for the most stable phase of each composition of $P_xC_{1-x}$ are displayed in Fig. 5(d). The absorption spectrum of graphene and black-phosphorene are in good agreement with the available theoretical results by PBE [45,46]. It is clear that the absorption of $P_xC_{1-x}$ is mainly dominated in the far-ultraviolet (FUV) region, similar to that of black-phosphorene, except for $P_{0.125}C_{0.875}$ which exhibits absorbance of ~10% over a broad energy range from the extreme-ultraviolet (EUV) region to FUV. It is interesting to note that the absorbance of $P_xC_{1-x}$ increases with composition and reaches 36% at $x = 0.875$ which is comparable to the absorbance of graphene.

## IV. CONCLUSION

The temperature-composition phase diagram is obtained for monolayer $P_xC_{1-x}$ using the first-principles method in combination with cluster expansion and Monte Carlo simulation. It is found that monolayer $P_xC_{1-x}$ energetically favours phase separation, rather than forming solid solutions, suggesting difficulty in growing monolayer $P_xC_{1-x}$ alloys using equilibrium growth method. Considering that monolayer $P_xC_{1-x}$ could be grown using non-equilibrium methods, we carried out a systematic bottom-up design and high-throughput first-principles calculations, to determine the structures of $P_xC_{1-x}$ for $x = 0.125, 0.25, 0.375, 0.5, 0.625, 0.75$ and $0.875$, and predict their properties. The most stable phases of monolayer $P_xC_{1-x}$ are of superlattice structures consisting of alternating graphene and black phosphorene nanoribbons along the armchair direction. The calculated mechanical properties, phonon spectra and *ab initio* molecular dynamics confirm that these structures, once grown, are mechanically, dynamically and thermodynamically stable. Monolayer $P_xC_{1-x}$ for all compositions considered are semiconductors with PBE bandgap ranging from 0.2 eV to 1.2 eV. High carrier mobilities and high optical absorption in the extreme UV and far UV regions are predicted, suggesting promising applications of the 2D PCs in electronics and optoelectronics.

# ACKNOWLEDGMENTS

This research project is supported by the Ministry of Education, Singapore, under its MOE AcRF Tier 2 Award MOE2019-T2-2-30 and AcRF Tier 1 Award WBS R-144-000-413-114. The authors acknowledge the Centre for Advanced 2D Materials, National University of Singapore, and the National Supercomputing Centre of Singapore for providing computational resources.

# STRUCTURAL DATA

The structures and relevant information of the monolayer $P_xC_{1-x}$ will be made available in the 2D materials database, in 2DMatPedia, located at http://www.2dmatpedia.org/.